\documentclass[sigconf,natbib=true,anonymous=false]{acmart}

\AtBeginDocument{%
  \providecommand\BibTeX{{%
    \normalfont B\kern-0.5em{\scshape i\kern-0.25em b}\kern-0.8em\TeX}}}

\setcopyright{acmcopyright}
\copyrightyear{2023}
\acmYear{2023}
\acmDOI{XXXXXXX.XXXXXXX}

\acmConference[SIGIR '23]{Proceedings of the 46th International ACM SIGIR Conference on Research and Development in Information Retrieval}{July 23--27, 2023}{Taipei, Taiwan}
\acmPrice{15.00}
\acmISBN{978-1-4503-XXXX-X/18/06}

\usepackage{adjustbox}
\usepackage{subfigure}

\begin{document}

\title{A Pairwise Dataset for GUI Conversion and Retrieval between Android Phones and Tablets}

% \affiliation{%

\author{Han Hu}
\affiliation{%
  \institution{Monash University}
  \city{Melbourne}
  \country{Australia}
}
\email{han.hu@monash.edu}

\author{Haolan Zhan}
\affiliation{%
  \institution{Monash University}
  \city{Melbourne}
  \country{Australia}
}
\email{haolan.zhan@monash.edu}

\author{Yujin Huang}
\affiliation{%
  \institution{Monash University}
  \city{Melbourne}
  \country{Australia}
}
\email{yujin.huang@monash.edu}

\author{Di Liu}
\affiliation{%
  \institution{Monash University}
  \city{Melbourne}
  \country{Australia}
}
\email{dliu0024@student.monash.edu}

\renewcommand{\shortauthors}{Hu and Zhan, et al.}
\newcommand{\han}[1]{\textcolor{red}{#1}}
\begin{abstract}
With the popularity of smartphones and tablets, users have become accustomed to using different devices for different tasks, such as using their phones to play games and tablets to watch movies.
To conquer the market, one app is often available on both smartphones and tablets.
However, although one app has similar graphic user interfaces (GUIs) and functionalities on phone and tablet, current app developers typically start from scratch when developing a tablet-compatible version of their app, which drives up development costs and wastes existing design resources.
Researchers are attempting to employ deep learning in automated GUIs development to enhance developers' productivity.
Deep learning models rely heavily on high-quality datasets.
There are currently several publicly accessible GUI page datasets for phones, but none for pairwise GUIs between phones and tablets.
This poses a significant barrier to the employment of deep learning in automated GUI development.
In this paper, we collect and make public the Papt dataset, which is a pairwise dataset for GUI conversion and retrieval between Android phones and tablets.

\end{abstract}






\maketitle

\section{Introduction}
Mobile apps are ubiquitous in our daily life for supporting different tasks such as reading, chatting, and banking.
Smartphones and tablets are the two types of portable devices with the most available apps~\cite{tabletShare}.
To conquer the market, one app is often available on both smartphones and tablets~\cite{majeed2015apps}.
Due to comparable functionalities, the smartphone and tablet versions of the same app have a highly similar Graphical User Interface (GUI).
Popular apps always share a similar GUI design between phone apps and tablet apps, for example, YouTube~\cite{youtube} and Spotify~\cite{spotify}.
If a tool can automatically recommend a GUI design for the appropriate tablet platform based on existing mobile GUIs, it can significantly minimise the developer's engineering effort and accelerate the development process.
From the user side, a comparable design could provide data or study on user preferences for consistency across different devices. 
It reduces the need for them to learn new navigation and interaction patterns~\cite{oulasvirta2020combinatorial}.
Therefore, automated GUI development tasks, such as the cross-platform conversion of GUI designs, GUI recommendations, etc., are gradually gaining attention from industry and academia~\cite{zhao2022code, li2022learning, chen2020wireframe}.
However, the field of automatic GUI development is still in a research bottleneck, lacking breakthroughs and widely recognized tools or methods.
If a present developer needs to develop a tablet-compatible version of their app, they usually start from scratch, resulting in needless costs increase and wasted existing design resources.

\begin{figure}[!t]%[htbp]
    \centering
    \includegraphics[width=\linewidth]{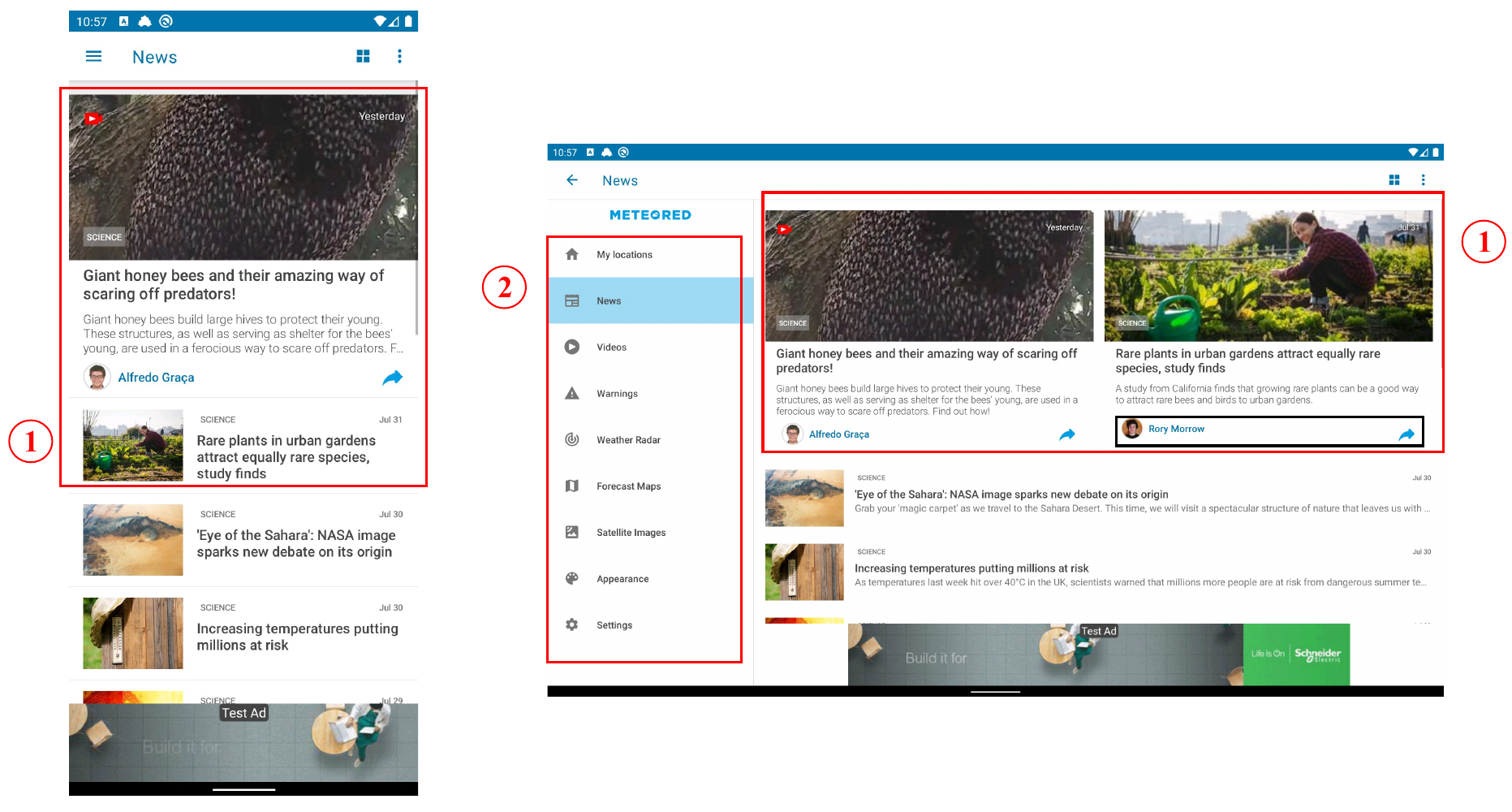}
    \caption{An example of a phone-tablet GUI pair of the app 'BBC News'. The GUI on the left is from the phone, while the GUI on the right is from the tablet.}
    \label{fig:introExp}
    \vspace{-0.3cm}
\end{figure}

According to our observations, the growth of automated GUI development is hindered by two reasons.
First, as deep learning approaches, particularly generative models~\cite{theis2015note}, become more widespread in the field of automated GUI development, researchers increasingly need a pairwise, high-quality GUI dataset for training models, summarizing rules, and so on.
The current datasets, for example, Rico~\cite{deka2017rico}, ReDraw~\cite{moran2019redraw}, and Clay~\cite{li2022learning}, only include single GUI pages with UI metadata and UI screenshots, and there are no pairwise corresponding GUI page pairs.
Current datasets are only suitable for GUI component identification, GUI information summarising and GUI completion.
The lack of valid GUI pairs in current datasets have severely hindered the growth of GUI automated development.
Second, the collecting of GUI pairs between phones and tablets is more labor-intensive.
Individual GUI pages can be automatically collected and labeled by current GUI testing and exploration tools~\cite{hu2011automating, memon2002gui} to speed up the collection process.
However, due to the disparity in screen size and GUI design between phones and tablets, it is challenging to automatically align the content on both GUI pages.
Figure~\ref{fig:introExp} shows an example of a phone-tablet GUI pair of the app 'BBC News'~\cite{bbcNews}.
To accommodate tablet devices, the UI component group 1 in the phone GUI are converted to the parts marked as 1 in the tablet GUI.
We can find that these components not only change their positions and sizes, but also the layouts and types of GUIs.
To keep a consistent layout with the left GUI components in the tablet GUI, the UI component group 1 in the tablet design adds new contents that are not available in the mobile GUI (black box in the tablet GUI).
Another UI component group, which is marked as 2 in the tablet GUI, is not present in the phone GUI at all.
One tablet GUI page may correspond to the contents of multiple phone GUI pages due to the different screen size and design style.
The UI contents in group 2 of the tablet GUI correspond to other phone GUIs.

In response to above challenges, we provide our dataset Papt, which is a \textbf{PA}irwise dataset for GUI conversion and retrieval between Android \textbf{P}hones and \textbf{T}ables.
It is the first pairwise GUI dataset of phones and tablets. 
The dataset contains 1,0035 corresponding phone-table GUI pairs, which are collected from 5,593 tablet-phone app pairs.
We first describe the data source, dataset collection process, collection approaches, and collection tools in this paper. 
We also open source our data collection tool for further related works.
Then, we describe the format of the dataset and the loading method. 
% Finally, we perform preliminary experiments on the dataset for the GUI conversion task and GUI retrieval task. 
% We present some of the GUI pages generated by the preliminary experiments.
% We discuss the challenges faced by the current models on GUI conversion and GUI retrieval tasks.
This resource paper bridges the gap between smartphone GUIs and tablet GUIs.
Our goal is to provide an effective benchmark for GUI automation development and to encourage more academics to explore this field.

In summary, the contributions of this paper are the following:
\begin{itemize}
    \item We contribute the first pairwise GUI dataset between Android phones and tablets~\footnote{\url{https://github.com/huhanGitHub/papt}}.

    \item We provide the detailed procedure of data collection and open source our data collection tool.

\end{itemize}
\section{Related Work}
% \han{(1) refer to the related work of the paper: https://arxiv.org/abs/2104.02416.}

% \han{(2) introduce existing Android GUI datasets. refer to the relate work of rico paper: https://dl.acm.org/doi/10.1145/3126594.3126651. refer to Chi2022 paper: https://arxiv.org/pdf/2201.04100.pdf}

\subsection{Android GUI dataset}
The research community has collected datasets to facilitate diverse deep learning-based applications in the realm of mobile app design.
Rico~\cite{deka2017rico} is the largest publicly available Android GUI dataset containing 72,219 screenshots from 9,772 apps.
It was built by the combination of crowdsourcing and automation and has been widely used as the primary data source for GUI modeling research.
However, several weaknesses~\cite{deka2021early,lee2020guicomp}, such as noise and erroneous labeling, have been identified in Rico.
To tackle these issues, a series of Android GUI datasets curated utilizing filtering and repairing mechanisms are introduced.
Enrico~\cite{leiva2020enrico} is the first enhanced dataset drawn from Rico, which is used for mobile layout design categorization and consists of 1460 high-quality screenshots produced by human revision, covering 20 different GUI design topics.
The VINS dataset~\cite{bunian2021vins}, developed specifically for detecting GUI elements, comprises 2,740 Android screenshots manually collected from various sources including Rico and Google Play.
Since the data cleaning process for both the Enrico and VINS datasets involves humans, adopting such approaches to improve existing Android GUI datasets at scale is expensive and time-consuming.
To this end, CLAY~\cite{li2022learning} employs deep learning models to automatically denoise Android screen layouts and create a large-scale GUI dataset with 59,555 screenshots on the basis of Rico. 
Apart from these Rico-based GUI datasets, several works~\cite{chen2018ui,chen2019gallery,chen2020wireframe,chen2020unblind,wang2021screen2words, hu2019code, hu2023automated, hu2023look, hu2023pairwise, hu2023first, chen2021my, huang2021robustness} also build their datasets for various GUI-related tasks such as Skeleton Generation, search and component prediction.

Despite improvements in recent Android GUI datasets, they are lack of updates so that some of their GUI styles are out-of-date.
More importantly, none of the datasets provides pairwise GUI pages between different mobile devices.
To fill the current gaps, we first introduce a pairwise dataset consisting of 10,035 phone-tablet GUI page pairs collected from 5,593 phone-tablet app pairs~\cite{hu2023pairwise}, which can be used for GUI conversion, retrieval, and recommendation between Android phones and tablets.

\subsection{Layout Generation}
As one of the main goals for our dataset is to facilitate the GUI conversion between Android and tablet apps, we apply layout generation techniques to our dataset.
Here we present a succinct review of existing layout generation techniques.
LayoutGAN~\cite{li2019layoutgan} is the first approach that utilizes a generative model (i.e., Generative Adversarial Network) to generate layouts.
In particular, it adopts self-attention layers to generate a realistic layout and proposes a novel differentiable wireframe rendering layer to enable Convolutional Neural Network (CNN)-based discrimination.
LayoutVAE~\cite{jyothi2019layoutvae} is an autoregressive generative model, which uses Long Short-Term Memory~\cite{hochreiter1997long} to consolidate the information from multiple UI elements and leverages Variational Autoencoders (VAEs)~\cite{kingma2013auto} to generate layouts.
Recently, VTN~\cite{arroyo2021variational} also exploits VAE architecture but both encoder and decoder are substituted with Transformers~\cite{vaswani2017attention}.
Equipped with self-attention layers, VTN possesses the capacity to learn appropriate layout arrangements without annotations.
At the same time, LayoutTransformer~\cite{gupta2021layouttransformer}, a purely Transformer-based framework, is proposed for layout generation.
It captures co-occurrences and implicit relationships among elements in layouts and uses such captured features to produce layouts with bounding boxes as units.
Based on our experiment results, we find that current layout generation models have limited capacity for GUI conversion between Android and tablet layouts, suggesting the potential research direction in automated GUI development.

\subsection{GUI Design Search}
Another goal of our dataset is to support the GUI retrieval that searches and recommends the comparable tablet GUI design in accordance with an Android GUI design.
We thus employ GUI design search techniques for our dataset.
Many research efforts have been made in GUI design search in recent years.
Rico~\cite{deka2017rico} is a neural-based training framework that aims to facilitate query-by-example search.
It provides a layout-encoding vector representation for each UI and offers a variety of visual representations for search engines.
GUIFetch~\cite{behrang2018guifetch} searches GUI design by leveraging a code-search technique, which retrieves the most similar GUI code for users based on their provided sketches.
WAE~\cite{chen2020wireframe} is a wireframe-based searching model that utilizes image autoencoder architecture to address the challenge of labeling large-scale GUI designs.
In light of the experiment results, deep learning-based approaches are able to achieve satisfactory performance.
Moreover, we anticipate the research community can dedicate more effort to the GUI retrieval between Android and tablet apps as this area is yet under-explored.

\section{Dataset}
In this section, we introduce the data source of this dataset in subsection~\ref{sec:datasource}, the collection approach in subsection~\ref{sec:collectApproach}, the collection tool in subsection~\ref{sec:tool}, the format of a GUI pair in subsection~\ref{sec:pairFormat}, the statistics of the dataset in subsection~\ref{sec:statistics}, and how to access the dataset in subsection~\ref{sec:accessDb}.
We also illustrate the advantages of our dataset compared to other current datasets in subsection~\ref{sec:dataCompare}.

\begin{table}[htbp]
\setlength{\abovecaptionskip}{5pt} 
\caption{The top 15 categories of data source apps.}
\begin{tabular}{ccc}
\hline
\textbf{Category}           & \textbf{\#Count} & \textbf{Percentage (\%)} \\
\hline
Entertainment      & 496     & 8.87         \\
Social             & 394     & 7.04         \\
Communication      & 326     & 5.83         \\
Lifestyle          & 318     & 5.69         \\
Books \& Reference & 286     & 5.11         \\
Education          & 279     & 4.98         \\
News \& Magazines  & 271     & 4.85         \\
Shopping           & 270     & 4.83         \\
Sports             & 267     & 4.78         \\
Music \& Audio     & 266     & 4.76         \\
Weather            & 265     & 4.73         \\
Finance            & 262     & 4.68         \\
Bussiness          & 261     & 4.67         \\
Travel \& Local    & 255     & 4.57         \\
Medical            & 254     & 4.54        \\
\bottomrule
\end{tabular}
\label{tab:appSource}
\vspace{-0.3cm}
\end{table}

\subsection{Data Source}
\label{sec:datasource}
We first crawl 6,456 tablet apps from Google Play.
Then we match their corresponding phone apps by their app names and app developers.
Finally, we collect 5,593 valid phone-tablet app pairs from 22 app categories.
Table~\ref{tab:appSource} shows the top 15 categories of 5,593 app pairs.
% Due to the effect of the data's long tail disctribution, we only display the top 15 categories.
The column \emph{Category} represents the category of these apps.
The column \emph{\#Count} and \emph{Percentage(\%)} denote the number of apps in this category and their percentage of the overall number of apps, respectively.
These 5,593 phone-tablet app pairs are the data source for this dataset.
The three most common categories of apps in the data source are: \emph{Entertainment} (8.87\%), \emph{Social} (7.04\%) and \emph{Communication} (5.83\%).
As shown in Table~\ref{tab:appSource}, the categories of apps in our data source are scattered and balanced.
Most of the categories occupy between 4\% and 6\% of the total dataset.
This balanced distribution ensures the dataset's generalizability and diversity.

\subsection{Data Collection Procedure}
\label{sec:collectApproach}
During the data collection process, we collect data in two stages: performing algorithms to automatically pair phone-tablet GUI pages and manually validating collected pairs.

In this section, we first introduce the current GUI data format in subsection~\ref{sec:GUIformat}.
Second, we illustrate two ways we use to match GUI pairs: by dynamically adjusting the resolution of the device (illustrated in subsection~\ref{sec:resolution} ) and by calculating the similarities (illustrated in subsection~\ref{sec:GUISim}).
We then develop data collecting tools based on these methodologies and use them to automatically collect GUI pairs.
After the automatic collection, three volunteers with at least one year of GUI development experience will manually check and eliminate invalid pairs from the automatically collected pairs.

\subsubsection{GUI Data Format} 
\label{sec:GUIformat}
\begin{figure}[!t]
    \centering
    \includegraphics[width=0.95\linewidth]{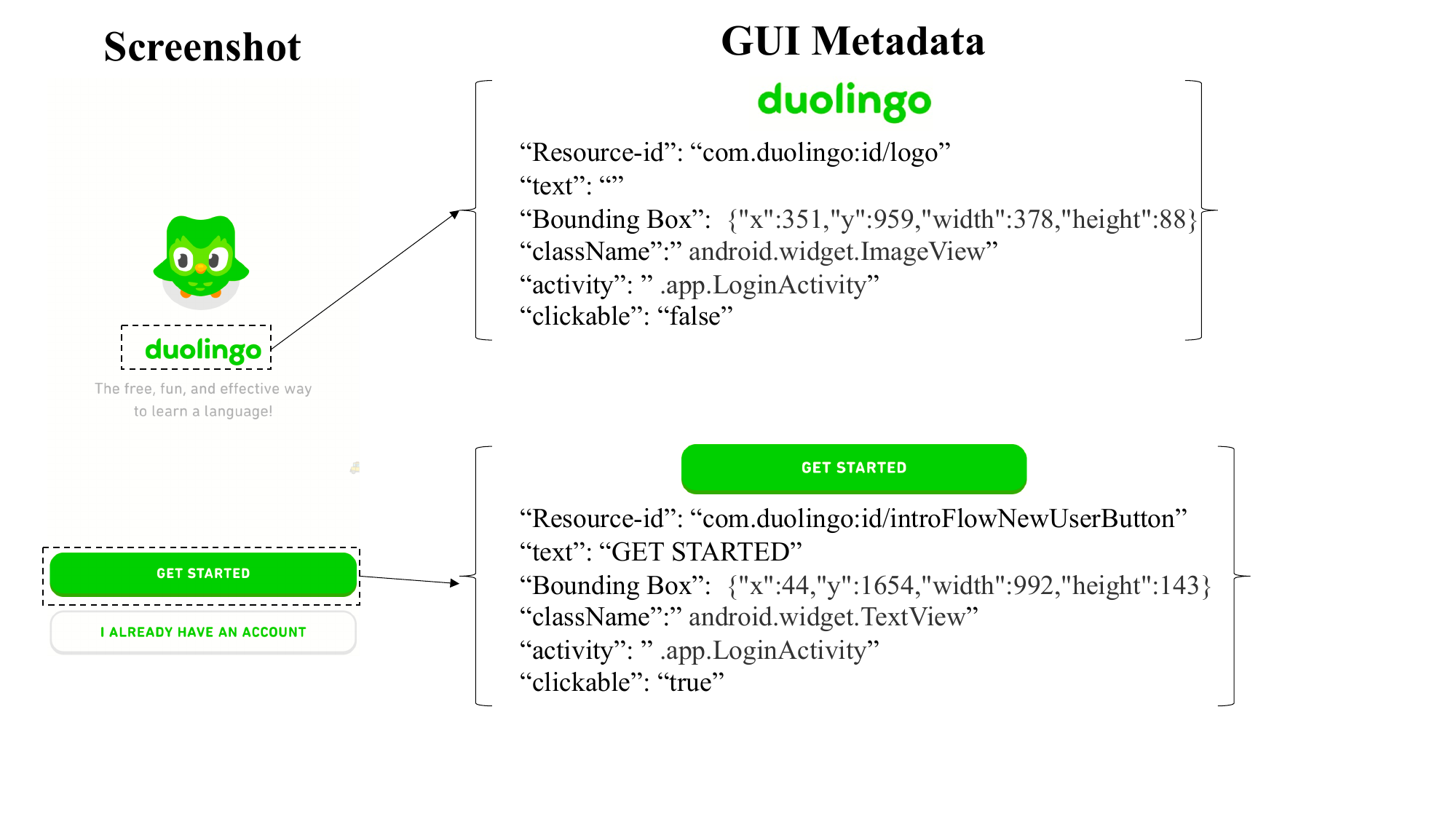}
    \caption{A screenshot of a GUI and metadata of some UI components inside the GUI. This example is from the app 'duolingo'.}
    \label{fig:metaExp}
    \vspace{-0.3cm}
\end{figure}

We install and run phone-tablet app pairs on the Pixel6 and Samsung Galaxy tab S8, respectively.
We use uiautomator2~\cite{uiautomator2} to collect screenshots and GUI metadata of the dynamically running apps.
Figure~\ref{fig:metaExp} shows an example of a collected GUI screenshot and metadata of some UI components inside the GUI.
This example is from the app 'duolingo'~\cite{duolingo}.
The metadata is a documentary object model (DOM) tree of current GUIs, which includes the hierarchy and properties (e.g., class, bounding box, layout) of UI components. 
We can infer the GUI hierarchy from the DOM tree hierarchy in metadata.

\subsubsection{Match GUI Pairs by Dynamically Adjusting the Device Resolution}
\label{sec:resolution}

Android's Responsive/Adaptive layouts provide an optimized user experience regardless of screen size, allowing Android apps to support phones, tablets, foldable, ChromeOS devices, portrait/landscape orientations, and resizable configurations~\cite{adpLay}.
Therefore, some apps define layout files that support different devices and call the corresponding layout file depending on the resolution of the installed device. 

These apps share the same APK files on both mobile and tablet devices.
We first deal with this type of apps.
According to Android's official guidelines for supporting different screen sizes~\cite{supportDifScr}, Android developers can provide alternate layouts for displays with a minimum width measured in density-independent pixels (dp or dip) by using the smallest width screen size qualifiers.
For example, developers can define two layout files for the \emph{MainActivity}: \emph{res/layout/main\_activity.xml} and \emph{res/layout-sw600dp/main\_activity.xml} for smartphones and tablets with 600 density-independent pixels, respectively.
The smallest width qualifier (sw600dp) specifies the smallest of the screen's two sides, regardless of the device's current orientation. 
This layout file allows us to determine that the current app's \emph{MainActivity} is optimal for the tablet's layout.
Two types of the smallest width qualifier: 600dp and 720dp specifically develop for the 7'' and 10'' tablets currently on the market~\cite{supportDifScr}.

\begin{figure}[htbp]
    \centering
    \includegraphics[width=\linewidth]{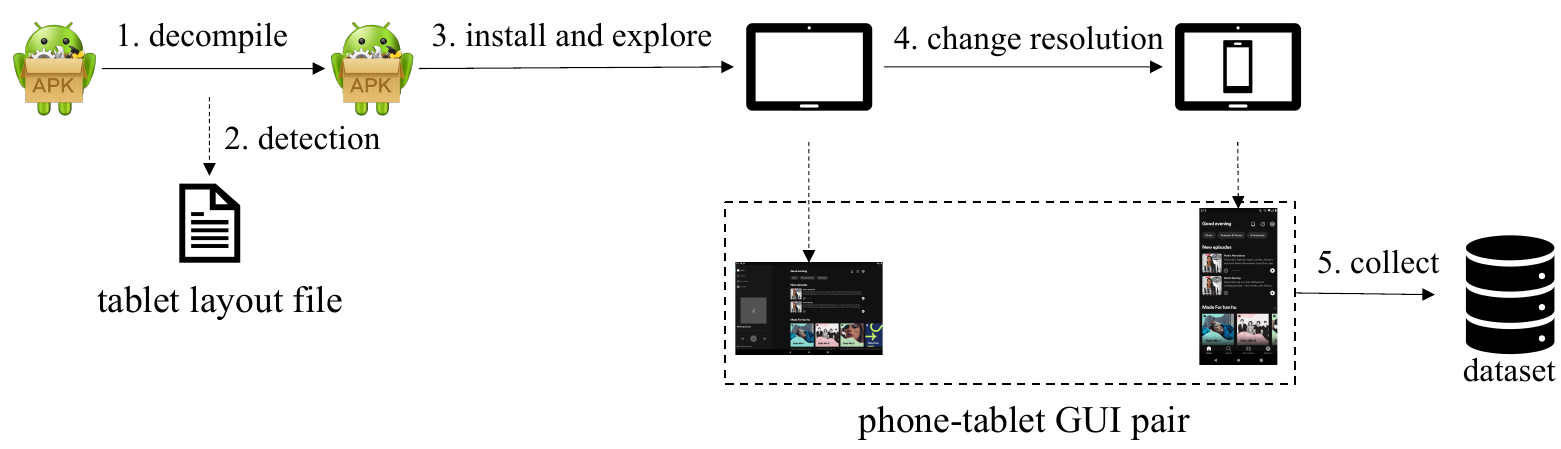}
    \caption{Pipeline of matching GUI pairs by dynamically adjusting the device resolution.}
    \label{fig:alg1}
    \vspace{-0.3cm}
\end{figure}

Figure~\ref{fig:alg1} shows the pipeline of matching GUI pairs by dynamically adjusting device resolutions.
Based on the above analysis, we first decompile the collected app pairs and then search for the presence of layout files for 600dp tablets and 720dp tablets in the source files (step 1 \& 2 in Figure~\ref{fig:alg1}).
In 5,593 app pairs, 1,214 pairs are found in the layout files of the smallest width qualifiers for the tablet device and share the same app on their phones and tablets.
Therefore, these 1,214 pairs use Responsive/Adaptive layouts to dynamically change GUI layout based on the installed device.

For these app pairs, we use the Windows Manager Command of ADB (\emph{adb shell wm size})~\cite{adb} to dynamically adjust the device's resolution.
We sequentially simulate the tablet and phone resolutions on a tablet device and allow the GUIs collected before and after the simulations to automatically pair up (steps 4 \& 5 in Figure~\ref{fig:alg2}).

\subsubsection{Match GUI Pairs by Comparing UI Similarities}
\label{sec:GUISim}

\begin{figure}[!t]
    \centering
    \includegraphics[width=\linewidth]{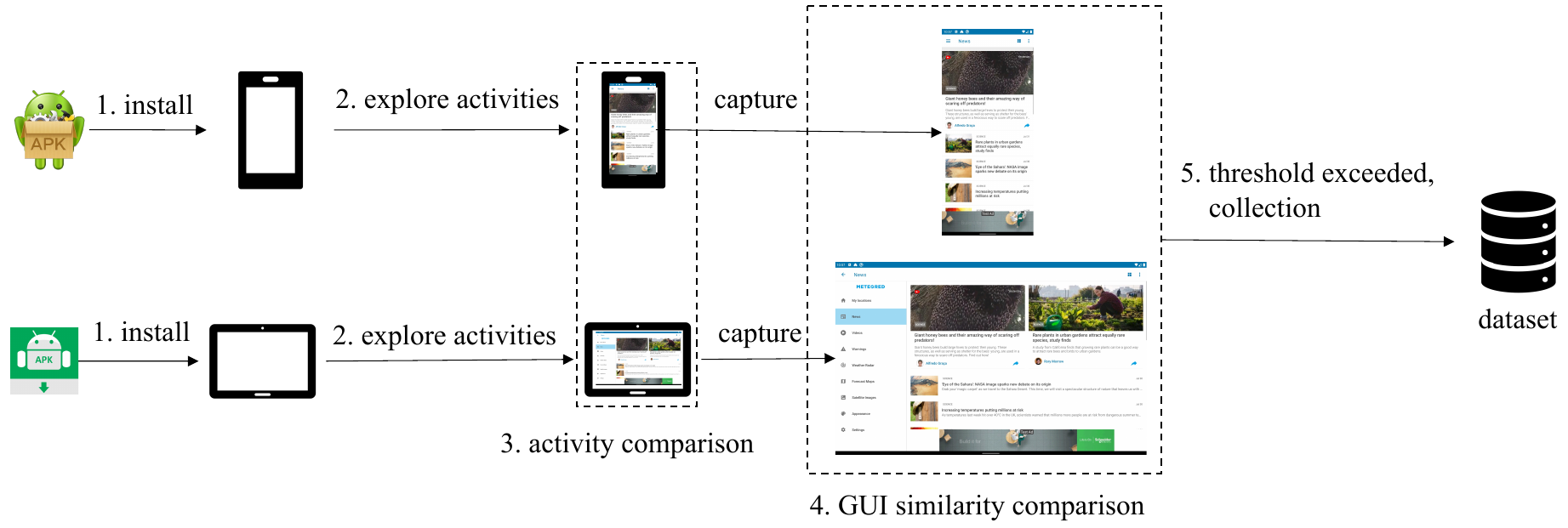}
    \caption{Pipeline of matching GUI pairs by comparing UI similarities.}
    \label{fig:alg2}
    \vspace{-0.3cm}
\end{figure}

With the exception of 1,214 apps that share the same file on tablet and phone, the remaining 4,379 phone-tablet pairs all have independent app files.
After dynamically launching the mobile and tablet apps, we match the corresponding GUI page pairs by comparing the similarity of their GUI pages.

Figure~\ref{fig:alg2} shows the pipeline of matching GUI pairs by comparing UI similarities.
Each app has many GUI pages, and a simple comparison of all pages between the mobile app and the tablet app would be time-consuming.
An Android activity provides the window in which the app draws its UI~\cite{activity}.
We first find the corresponding Android activity pairs of app pairs. 
Then we compare each GUI page on the basis of the corresponding Android activity pairs.
To match Android activity pairs, we extract activity names from each GUI and encode them into numerical semantic vectors using a pre-trained BERT~\cite{devlin2019bert} model.
We match the activity-level pairs by comparing their semantically close activity vectors  (Step 3 in Figure~\ref{fig:alg2}).
For example, the GUI in activity \textit{homeActivity} and \textit{mainActivity} are matched by close semantic vectors.
One Android activity may have multiple Android fragments~\cite{fragment} and GUI pages~\cite{machiry2013dynodroid, GUIstate} with different UI components and layouts in current industrial apps.
We compare the attributes of GUI components between phones and tablets to pair phone-tablet GUI pairs at lower granularity (Step 4 in Figure~\ref{fig:alg2}).
In pairing, UI components are identified by their types and properties.
UI components between phones and tablets  with the same types and properties are considered paired GUI components.
For example, two \emph{TextViews} with the same texts, two \emph{ImageViews} with the same images, two \emph{Buttons} with the same texts are considered the paired components.
If more than half of the UI components in two GUIs are paired, they are considered a phone-tablet GUI pair (Step 5 in Figure~\ref{fig:alg2}).

\subsubsection{Manual GUI Pair Verification}
\label{sec:manualPair}
Due to the limitations of ADB, current data collection tools cannot get the metadata of UI type \emph{WebView} and some user-defined third-party UI components.
Meanwhile, when some fragments and UIs of Android are covered, we only need the metadata of the UI at the front end (because the covered UI is not rendered and displayed on the current screen). However, the present data collection tools~\cite{adb, uiautomator2} capture information for both the viewable displayed and the covered UI, resulting in the collection of erroneous data.
Therefore, three volunteers with at least one year of Android development experience perform a second round of manual data validation.
Volunteers check each pair and then apply two criteria to evaluate the data quality: the data's reliability and the rationality of the pairs. 
Volunteers verify the validity of collected pairs and remove pairs with inaccurate metadata.
Volunteers also check the rationality of matched pairs and remove GUI pairs that do not correspond to each other.
In this process, our volunteers also manually match some phone-tablet GUI pairs.

\subsection{Data Collection Tool}
\label{sec:tool}
Based on the above-described two collection strategies, we develop two distinct collecting tools: the adjust resolution collector and the similarity matching collector.

The first tool dynamically adjusts the resolution of the current device using ADB instructions. 
When the running app detects a change in the screen's resolution, it will call the layout file designed for the tablet and change the layout of the current GUI.

The second tool concurrently runs two apps of one app pair on a mobile phone and a tablet.
The tool dynamically evaluates the similarity of the GUIs presented on two devices, and automatically collects the matched GUI page pair when the similarity exceeds a predetermined threshold.

These two data collection tools are also included in the repository of the publicly accessible dataset. 
With the installation instructions provided, more researchers can utilise our tools to collect more customised GUI datasets for future research.

\subsection{Format of A GUI Pair}
\label{sec:pairFormat}
In this section, we introduce the format of each GUI pair.

\begin{figure}[!t]
    \centering
    \includegraphics[width=0.9\linewidth]{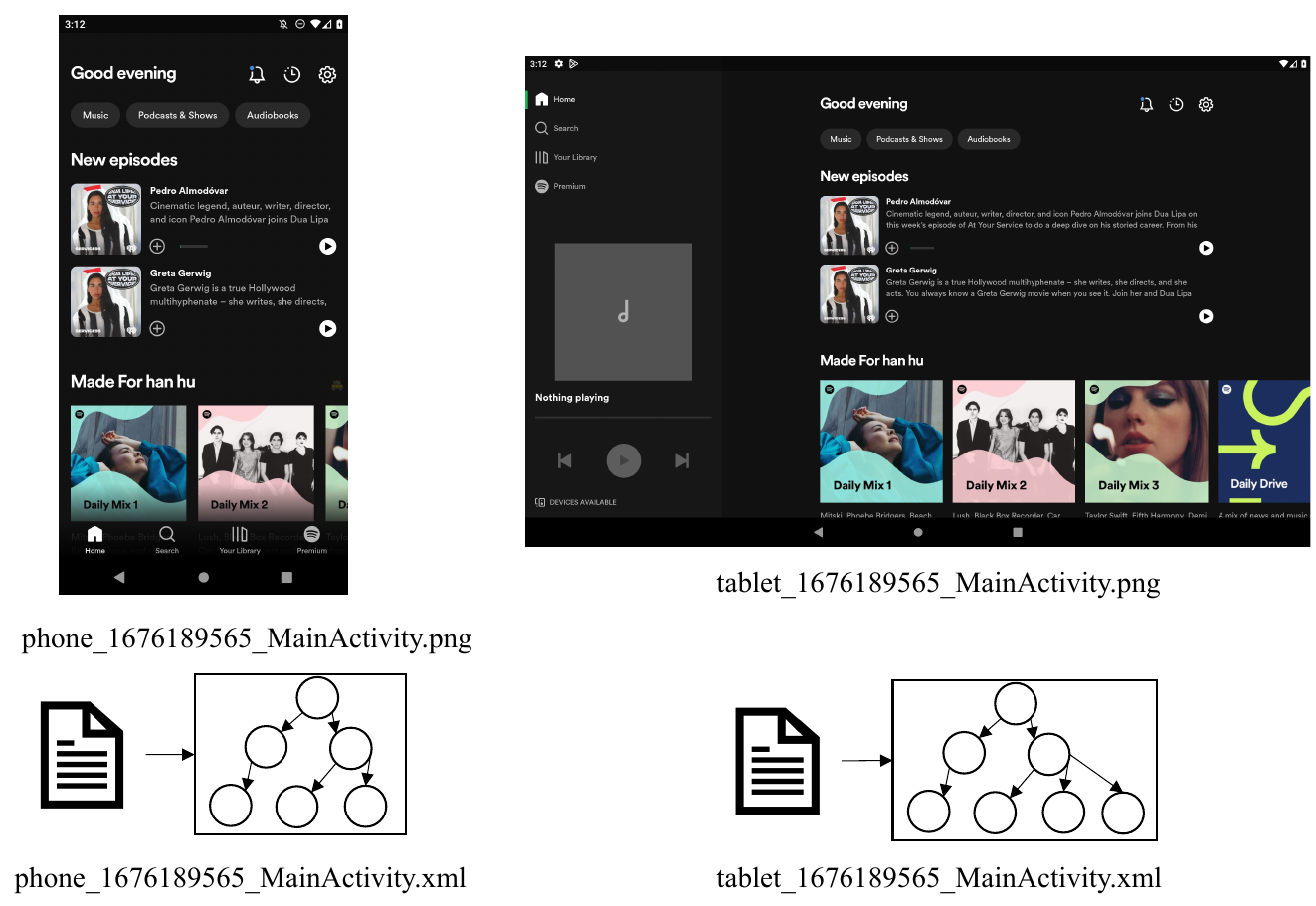}
    \caption{An example of paired GUI pages of 'Spotify'.}
    \label{fig:elem}
    \vspace{-0.3cm}
\end{figure}

Figure~\ref{fig:elem} shows an example of pairwise GUI pages of the app 'Spotify' in our dataset.
All GUI pairs in one phone-tablet app pair are placed in the same directory.
Each pair consists of four elements: a screenshot of the GUI running on the phone (\emph{phone\_1676189565\_-MainActivity.png}), the metadata data corresponding to the GUI screenshot on the phone (\emph{ phone\_1676189565\_MainActivity.xml
}), a screenshot of the GUI running on the tablet (\emph{tablet\_1676189565\_-MainActivity.png}
), and the metadata data corresponding to the GUI screenshot on the tablet (\emph{tablet\_1676189565\_MainActivity.xml
}).
The naming format for all files in the dataset is \emph{Device\_Timestamp\_Activity Name}.
As shown in Figure~\ref{fig:elem}, The filename \emph{tablet\_1676189565\_-MainActivity.xml} indicates that this file was obtained by the tablet and was collected with the timestamp \emph{1676189565}, this GUI belongs to \emph{MainActivity} and this file is a metadata file in XML format.
We use timestamps and activity names to distinguish phone-tablet GUI pairs.

\subsection{Statistics of the Database}
\label{sec:statistics}
We collect a total of 10,035 valid phone-tablet GUI pairs.
For a more comprehensive presentation of our dataset, we statistically analyze the collected pairs from two perspectives: the distribution of UI view types within the dataset in subsection~\ref{sec:disUI} and the distribution of GUI similarity between GUI pairs in subsection~\ref{sec:disSim}.

\subsubsection{Distribution of UI View Types}
\label{sec:disUI}
In Android development, a UI view is a basic building block for creating user interfaces. 
Views are responsible for drawing and handling user interactions for a portion of the screen~\cite{AndroidView}. 
For example, a button, a text , an image, and a list are all a type of view.

\begin{figure}[htbp]
    \centering
    \includegraphics[width=0.9\linewidth]{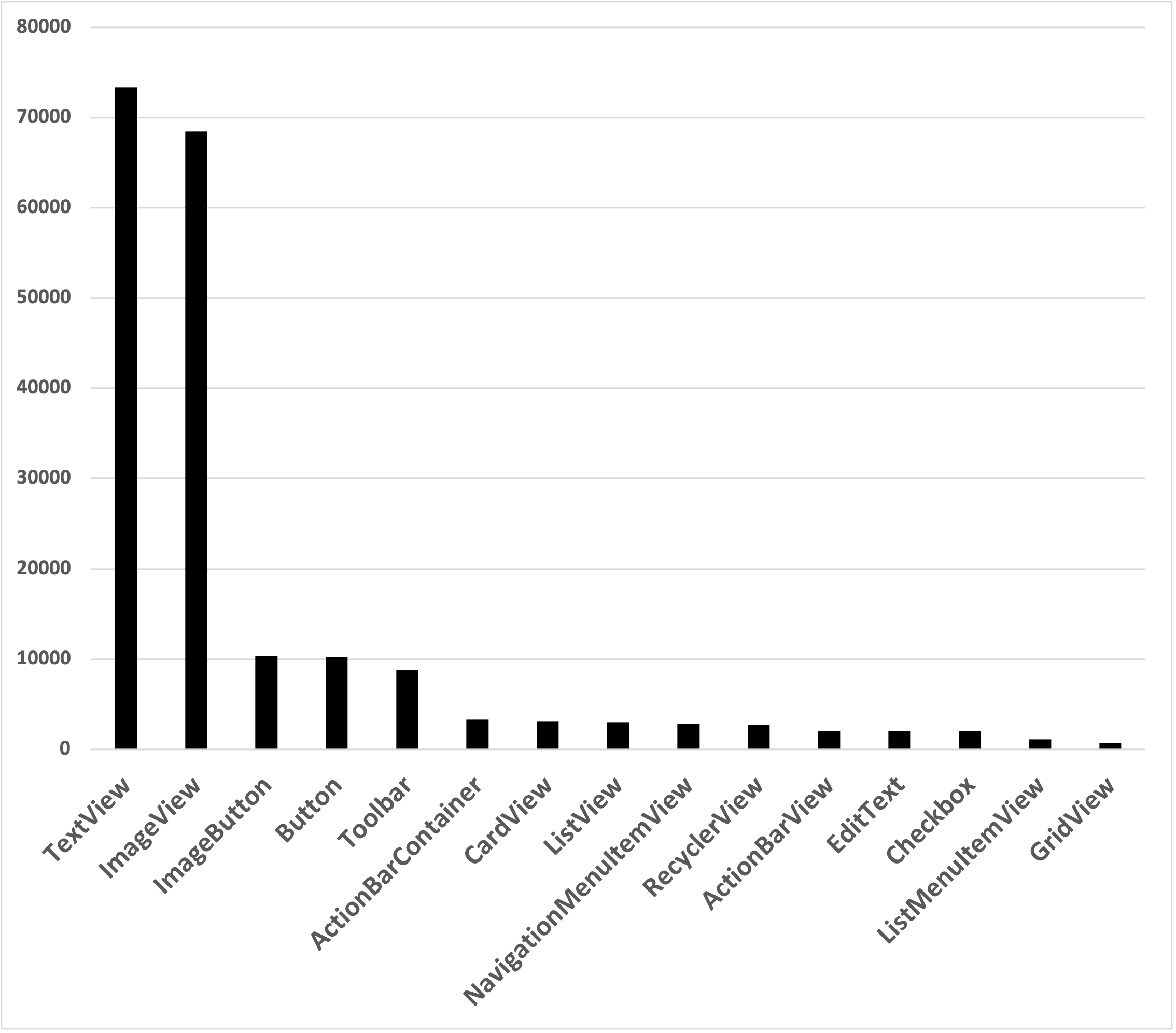}
    \caption{Distribution of top 15 UI view types in the dataset.}
    \label{fig:uiDis}
    \vspace{-0.3cm}
\end{figure}

Figure~\ref{fig:uiDis} illustrates the distribution of UI View types in the dataset.
Considering the data's long-tail distribution, we only display the top 15 types.
We can see that the \emph{TextView} and \emph{ImageView} types, including all their derived categories such as \emph{AppCompatTextView}, \emph{AppCompatCheckedTextView}, and \emph{AppCompatImageView}, are the most common UI view types in the dataset. 
Their numbers (73,349 and 68,496) significantly outnumber all other view types.
The GUI primarily presents information via text and images, so text and image-related views are the most prevalent in the database.
\emph{ImageButton} (10,366) and \emph{Button} (10,235) are the third and fourth most UI views.
Users interact with the GUI mainly through clicks and click operations rely heavily on button views, so button-related views are also common in GUI datasets.

\subsubsection{Distribution of GUI Pair Similarity}
\label{sec:disSim}
The similarity analysis between phone-tablet GUI pairs is important for downstream tasks.
Given a GUI pair, there are a total of $M$ and $N$ GUI views in the GUIs of the phone and tablet, respectively.
Suppose there are $L$ the same views in the GUIs of the phone and the tablet.
The similarity of their GUIs is calculated as
\begin{equation}
    Sim(M, N) = \frac{2 * L}{M+N}
\end{equation}

% In GUI pairs where the kind of UI views and their contents are the same, we presume that the two views are the same regardless of their size and position attributes.

\begin{figure}[htbp]
    \centering
    \includegraphics[width=0.9\linewidth]{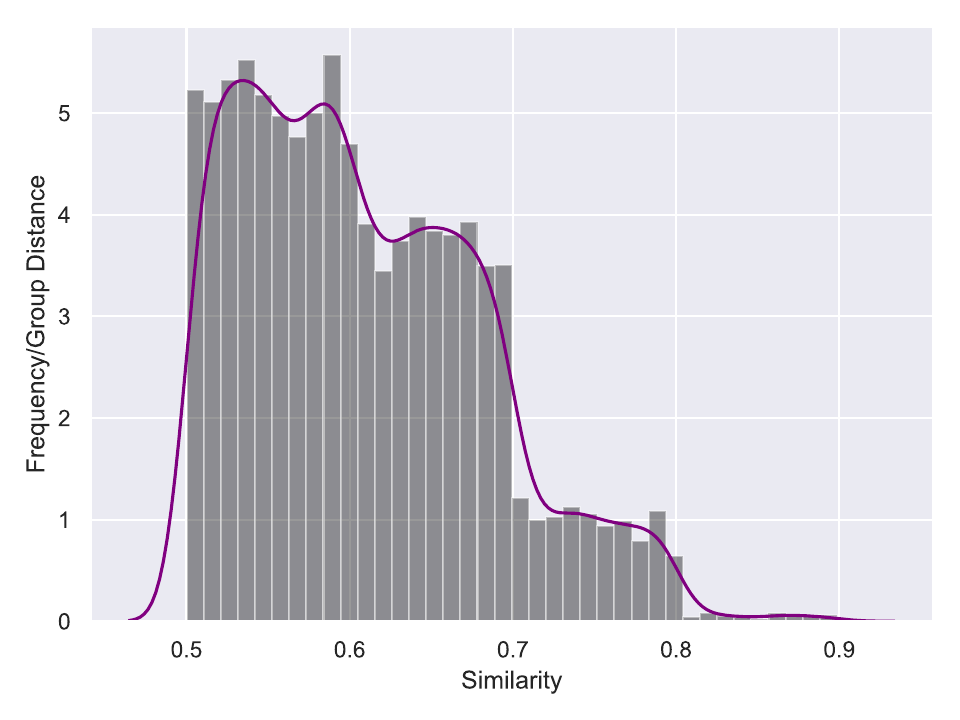}
    \caption{The frequency histogram of GUI similarity of collected pairs}
    \label{fig:simiDis}
\end{figure}

Figure~\ref{fig:simiDis} shows the frequency histogram of GUI similarities of our phone-tablet GUI pairs in the dataset.
The similarity between the GUIs of phone and tablet in most pairs is between 0.5 and 0.7.
Considering the difference in screen size between tablets and phones, the current phone GUI page can only contain part of the UI views in the corresponding tablet GUI page, and the current data reminds us that when performing downstream tasks such as GUI layout generation, search, etc., we should consider filling in the contents that are not available in the mobile phone GUI page.

\subsection{Accessing the Dataset}
\label{sec:accessDb}
The dataset is made accessible to the public in accordance with the criteria outlined in the attached license agreements\footnote{\url{https://github.com/huhanGitHub/papt}}.
The pairs in the dataset are contained in separate folders according to the app.
Most of the app folders are named after the package name of the app's APK, for example, \emph{air.com.myheritage.mobile} , and a few are named after the app's name, for example, \emph{Spotify}.
In each app folder, as described in Section~\ref{sec:pairFormat}, each pair contains four elements: the phone GUI screenshot, the XML file of the phone GUI metadata, the corresponding tablet GUI screenshot, and the XML file of the tablet GUI metadata.
% The current GUI page's activity name and a unique timestamp are used to identify different GUI page pairs.
We also shared the script for loading all GUI pairs in the open source repository.

\subsection{Compare With Available Datasets}
\label{sec:dataCompare}
\begin{table*}[htbp]
\setlength{\abovecaptionskip}{0pt} 
\setlength{\belowcaptionskip}{5pt}
\caption{Comparison between our dataset and other GUI datasets}
\begin{adjustbox}{ width=\textwidth,center}
\centering
\begin{tabular}{ccccccc}
\hline
\textbf{Dataset}   & \textbf{GUI Platform} & \textbf{\#GUIs} & \textbf{\#Paired GUIs} & \textbf{\#Data Source App} & \textbf{Latest Updates} & \textbf{Mainly targeted tasks}\\
\hline
Rico~\cite{deka2017rico} & Phone & ~72,000 & 0 & ~9,700 & Sep. 2017 & UI Component Recognition, GUI completion \\
UI2code~\cite{chen2018ui} & Phone & ~185,277 & 0 & ~5,043 & June. 2018 & UI Skeleton Generation\\
Gallery D.C.~\cite{chen2019gallery} & Phone & 68,702 & 0 & 5,043 & Nov. 2019 & UI Search \\
LabelDroid~\cite{chen2020unblind} & Phone & 394,489 & 0 & 15,087 & May. 2020 & UI Component Prediction \\
UI5K~\cite{chen2020wireframe} & Phone & 54,987 & 0 & 7,748 & June. 2020 & UI Search \\
Enrico~\cite{leiva2020enrico} & Phone & ~1,460 & 0 & ~9,700 & Oct. 2020 & UI Layout Design Categorization \\
VINS~\cite{bunian2021vins} & Phone & ~2,740 & 0 & ~9,700 & May. 2021 & UI Search \\
Screen2Words~\cite{wang2021screen2words} & Phone & 22,417 & 0 & 6,269 & Oct. 2021 & UI screen summarization \\
Clay~\cite{li2022learning} & Phone & 59,555 & 0 & ~9,700 & May. 2022 &  UI Component Recognition, GUI completion \\
\textbf{Papt} & Phone, Tablet & 20,070 & 10,035 & 11,186 & Jan. 2023 & UI Component Recognition, GUI completion, GUI conversion, GUI search \\ 

\bottomrule
\end{tabular}
\end{adjustbox}
\label{tab:compare}
\vspace{-0.3cm}
\end{table*}

\subsubsection{Application to More Tasks}
Table~\ref{tab:compare} shows a summary of our and other GUI datasets.
First, since our data consist of phone-tablet pairs, we must manually locate the corresponding GUI pages between phones and tablets, resulting in a lesser number of pages than comparable datasets.
However, we now have a broader data source (including tablet GUIs), more supported tasks, and newer data.
Notably, it is the only available GUI dataset that contains phone-tablet pairwise GUIs. 
Our dataset addresses numerous significant gaps in existing GUI automated development and provides effective data support for the application of deep learning techniques in GUI generation, search, recommendation, and other domains.

\subsubsection{Data Accuracy}
\begin{figure}[htbp]
    \centering
    \includegraphics[width=\linewidth]{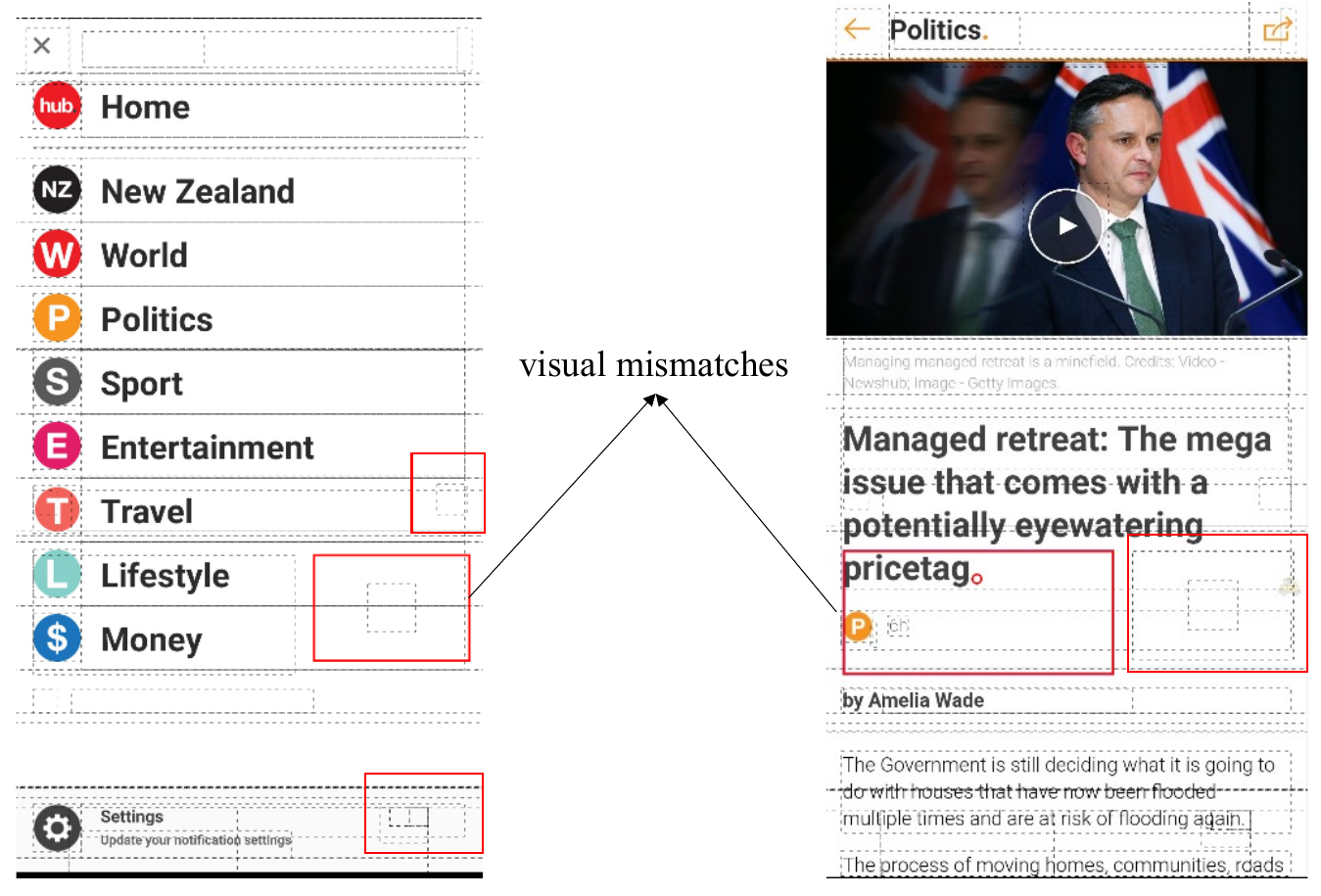}
    \caption{Examples of visual mismatches in current GUI datasets.}
    \label{fig:mismatch}
    \vspace{-0.3cm}
\end{figure}

Specifically, our data eliminates a large number of GUI visual mismatches that are frequent in current datasets like as Rico and Enrico.
Due to the limitations of the previous data collection tools, some GUIs have visual mismatches in the metadata and screenshots.
Figure~\ref{fig:mismatch} shows typical visual mismatch examples between hierarchy metadata and screenshots in current datasets.
Based on the bounding box coordinates of Android views provided in the metadata, we depict the location of the views in the metadata as a black dashed line in the screenshot.
We mark the obvious visual mismatches with a solid red line box, which do not correspond to any of the views in the rendered screenshot.
The metadata provides information on the UI elements behind the current layer, but these elements cannot be interacted with on the current screenshot.
Visual mismatches between the UI data in the metadata and the screenshot would result in the UI data in the metadata and the screenshot not corresponding one to the other.
Too many mismatch cases would have a negative impact on the efficiency of model generation and search. 
The selected UI collecting tool, UIautomator2, has optimised the GUI caption technique to avoid metadata and screenshots from containing inconsistent UI information~\cite{uiautomator2}.
During manual reviews, our volunteers also eliminated GUI pages with mismatched.
Compared to other datasets, such as rico, fewer mismatches give us a higher accuracy of our data.

\section{Conclusion}
In this paper, we propose a pairwise dataset Papt for GUI conversion and retrieval between Android phones and tablets.
As far as we know, this is the first dataset aims to bridge phone and tablet GUIs.
We explain and analyze the data source of this dataset. 
We illustrate the two approaches for collecting pairwise GUIs and introduce our data collection tools.
We use an example to show the format of our GUI pairs and show the distribution of UI view types and GUI pair similarity in our dataset.
We also demonstrate how to access this dataset and compare the advantages of our dataset with current other GUI datasets.
% We perform preliminary experiments for two tasks: GUI conversion and GUI retrieval.
% We apply current widely used GUI generation and retrieval approaches in this dataset and analyze their performance.
% We show some examples of GUI pages generated by selected current approaches.
% We also discuss some valuable research problems based on current experimental results.

\bibliographystyle{ACM-Reference-Format}
\bibliography{sigir2023-ref}


\begin{thebibliography}{48}

%%% the Web address.

\ifx \showCODEN    \undefined \def \showCODEN     #1{\unskip}     \fi
\ifx \showDOI      \undefined \def \showDOI       #1{#1}\fi
\ifx \showISBNx    \undefined \def \showISBNx     #1{\unskip}     \fi
\ifx \showISBNxiii \undefined \def \showISBNxiii  #1{\unskip}     \fi
\ifx \showISSN     \undefined \def \showISSN      #1{\unskip}     \fi
\ifx \showLCCN     \undefined \def \showLCCN      #1{\unskip}     \fi
\ifx \shownote     \undefined \def \shownote      #1{#1}          \fi
\ifx \showarticletitle \undefined \def \showarticletitle #1{#1}   \fi
\ifx \showURL      \undefined \def \showURL       {\relax}        \fi
% invisible to TeX
\providecommand\bibfield[2]{#2}
\providecommand\bibinfo[2]{#2}
\providecommand\natexlab[1]{#1}
\providecommand\showeprint[2][]{arXiv:#2}

\bibitem[activity(2023)]%
        {activity}
activity \bibinfo{year}{2023}\natexlab{}.
\newblock \bibinfo{title}{activity}.
\newblock
\newblock
\newblock
\shownote{https://developer.android.com/reference/android/app/Activity}.


\bibitem[adb(2023)]%
        {adb}
adb \bibinfo{year}{2023}\natexlab{}.
\newblock \bibinfo{title}{adb}.
\newblock
\newblock
\newblock
\shownote{https://developer.android.com/studio/command-line/adb}.


\bibitem[adpLay(2023)]%
        {adpLay}
adpLay \bibinfo{year}{2023}\natexlab{}.
\newblock \bibinfo{title}{adpLay}.
\newblock
\newblock
\newblock
\shownote{https://developer.android.com/jetpack/compose/layouts/adaptive}.


\bibitem[AndroidView(2023)]%
        {AndroidView}
AndroidView \bibinfo{year}{2023}\natexlab{}.
\newblock \bibinfo{title}{AndroidView}.
\newblock
\newblock
\newblock
\shownote{https://data-flair.training/blogs/android-layout-and-views/}.


\bibitem[Arroyo et~al\mbox{.}(2021)]%
        {arroyo2021variational}
\bibfield{author}{\bibinfo{person}{Diego~Martin Arroyo}, \bibinfo{person}{Janis
  Postels}, {and} \bibinfo{person}{Federico Tombari}.}
  \bibinfo{year}{2021}\natexlab{}.
\newblock \showarticletitle{Variational transformer networks for layout
  generation}. In \bibinfo{booktitle}{\emph{Proceedings of the IEEE/CVF
  Conference on Computer Vision and Pattern Recognition}}.
  \bibinfo{pages}{13642--13652}.
\newblock


\bibitem[bbcNews(2023)]%
        {bbcNews}
bbcNews \bibinfo{year}{2023}\natexlab{}.
\newblock \bibinfo{title}{bbcNews}.
\newblock
\newblock
\newblock
\shownote{https://www.bbc.com/news}.


\bibitem[Behrang et~al\mbox{.}(2018)]%
        {behrang2018guifetch}
\bibfield{author}{\bibinfo{person}{Farnaz Behrang}, \bibinfo{person}{Steven~P
  Reiss}, {and} \bibinfo{person}{Alessandro Orso}.}
  \bibinfo{year}{2018}\natexlab{}.
\newblock \showarticletitle{GUIfetch: supporting app design and development
  through GUI search}. In \bibinfo{booktitle}{\emph{Proceedings of the 5th
  International Conference on Mobile Software Engineering and Systems}}.
  \bibinfo{pages}{236--246}.
\newblock


\bibitem[Bunian et~al\mbox{.}(2021)]%
        {bunian2021vins}
\bibfield{author}{\bibinfo{person}{Sara Bunian}, \bibinfo{person}{Kai Li},
  \bibinfo{person}{Chaima Jemmali}, \bibinfo{person}{Casper Harteveld},
  \bibinfo{person}{Yun Fu}, {and} \bibinfo{person}{Magy~Seif Seif El-Nasr}.}
  \bibinfo{year}{2021}\natexlab{}.
\newblock \showarticletitle{Vins: Visual search for mobile user interface
  design}. In \bibinfo{booktitle}{\emph{Proceedings of the 2021 CHI Conference
  on Human Factors in Computing Systems}}. \bibinfo{pages}{1--14}.
\newblock


\bibitem[Chen et~al\mbox{.}(2019)]%
        {chen2019gallery}
\bibfield{author}{\bibinfo{person}{Chunyang Chen}, \bibinfo{person}{Sidong
  Feng}, \bibinfo{person}{Zhenchang Xing}, \bibinfo{person}{Linda Liu},
  \bibinfo{person}{Shengdong Zhao}, {and} \bibinfo{person}{Jinshui Wang}.}
  \bibinfo{year}{2019}\natexlab{}.
\newblock \showarticletitle{Gallery dc: Design search and knowledge discovery
  through auto-created gui component gallery}.
\newblock \bibinfo{journal}{\emph{Proceedings of the ACM on Human-Computer
  Interaction}} \bibinfo{volume}{3}, \bibinfo{number}{CSCW}
  (\bibinfo{year}{2019}), \bibinfo{pages}{1--22}.
\newblock


\bibitem[Chen et~al\mbox{.}(2018)]%
        {chen2018ui}
\bibfield{author}{\bibinfo{person}{Chunyang Chen}, \bibinfo{person}{Ting Su},
  \bibinfo{person}{Guozhu Meng}, \bibinfo{person}{Zhenchang Xing}, {and}
  \bibinfo{person}{Yang Liu}.} \bibinfo{year}{2018}\natexlab{}.
\newblock \showarticletitle{From ui design image to gui skeleton: a neural
  machine translator to bootstrap mobile gui implementation}. In
  \bibinfo{booktitle}{\emph{Proceedings of the 40th International Conference on
  Software Engineering}}. \bibinfo{pages}{665--676}.
\newblock


\bibitem[Chen et~al\mbox{.}(2020a)]%
        {chen2020wireframe}
\bibfield{author}{\bibinfo{person}{Jieshan Chen}, \bibinfo{person}{Chunyang
  Chen}, \bibinfo{person}{Zhenchang Xing}, \bibinfo{person}{Xin Xia},
  \bibinfo{person}{Liming Zhu}, \bibinfo{person}{John Grundy}, {and}
  \bibinfo{person}{Jinshui Wang}.} \bibinfo{year}{2020}\natexlab{a}.
\newblock \showarticletitle{Wireframe-based UI design search through image
  autoencoder}.
\newblock \bibinfo{journal}{\emph{ACM Transactions on Software Engineering and
  Methodology (TOSEM)}} \bibinfo{volume}{29}, \bibinfo{number}{3}
  (\bibinfo{year}{2020}), \bibinfo{pages}{1--31}.
\newblock


\bibitem[Chen et~al\mbox{.}(2020b)]%
        {chen2020unblind}
\bibfield{author}{\bibinfo{person}{Jieshan Chen}, \bibinfo{person}{Chunyang
  Chen}, \bibinfo{person}{Zhenchang Xing}, \bibinfo{person}{Xiwei Xu},
  \bibinfo{person}{Liming Zhu}, \bibinfo{person}{Guoqiang Li}, {and}
  \bibinfo{person}{Jinshui Wang}.} \bibinfo{year}{2020}\natexlab{b}.
\newblock \showarticletitle{Unblind your apps: Predicting natural-language
  labels for mobile gui components by deep learning}. In
  \bibinfo{booktitle}{\emph{Proceedings of the ACM/IEEE 42nd International
  Conference on Software Engineering}}. \bibinfo{pages}{322--334}.
\newblock


\bibitem[Chen et~al\mbox{.}(2021)]%
        {chen2021my}
\bibfield{author}{\bibinfo{person}{Qiuyuan Chen}, \bibinfo{person}{Xin Xia},
  \bibinfo{person}{Han Hu}, \bibinfo{person}{David Lo}, {and}
  \bibinfo{person}{Shanping Li}.} \bibinfo{year}{2021}\natexlab{}.
\newblock \showarticletitle{Why my code summarization model does not work: Code
  comment improvement with category prediction}.
\newblock \bibinfo{journal}{\emph{ACM Transactions on Software Engineering and
  Methodology (TOSEM)}} \bibinfo{volume}{30}, \bibinfo{number}{2}
  (\bibinfo{year}{2021}), \bibinfo{pages}{1--29}.
\newblock


\bibitem[Deka et~al\mbox{.}(2021)]%
        {deka2021early}
\bibfield{author}{\bibinfo{person}{Biplab Deka}, \bibinfo{person}{Bardia
  Doosti}, \bibinfo{person}{Forrest Huang}, \bibinfo{person}{Chad Franzen},
  \bibinfo{person}{Joshua Hibschman}, \bibinfo{person}{Daniel Afergan},
  \bibinfo{person}{Yang Li}, \bibinfo{person}{Ranjitha Kumar},
  \bibinfo{person}{Tao Dong}, {and} \bibinfo{person}{Jeffrey Nichols}.}
  \bibinfo{year}{2021}\natexlab{}.
\newblock \showarticletitle{An Early Rico Retrospective: Three Years of Uses
  for a Mobile App Dataset}.
\newblock \bibinfo{journal}{\emph{Artificial Intelligence for Human Computer
  Interaction: A Modern Approach}} (\bibinfo{year}{2021}),
  \bibinfo{pages}{229--256}.
\newblock


\bibitem[Deka et~al\mbox{.}(2017)]%
        {deka2017rico}
\bibfield{author}{\bibinfo{person}{Biplab Deka}, \bibinfo{person}{Zifeng
  Huang}, \bibinfo{person}{Chad Franzen}, \bibinfo{person}{Joshua Hibschman},
  \bibinfo{person}{Daniel Afergan}, \bibinfo{person}{Yang Li},
  \bibinfo{person}{Jeffrey Nichols}, {and} \bibinfo{person}{Ranjitha Kumar}.}
  \bibinfo{year}{2017}\natexlab{}.
\newblock \showarticletitle{Rico: A mobile app dataset for building data-driven
  design applications}. In \bibinfo{booktitle}{\emph{Proceedings of the 30th
  annual ACM symposium on user interface software and technology}}.
  \bibinfo{pages}{845--854}.
\newblock


\bibitem[Devlin et~al\mbox{.}(2019)]%
        {devlin2019bert}
\bibfield{author}{\bibinfo{person}{Jacob Devlin}, \bibinfo{person}{Ming-Wei
  Chang}, \bibinfo{person}{Kenton Lee}, {and} \bibinfo{person}{Kristina
  Toutanova}.} \bibinfo{year}{2019}\natexlab{}.
\newblock \bibinfo{title}{BERT: Pre-training of Deep Bidirectional Transformers
  for Language Understanding}.
\newblock
\newblock
\showeprint[arxiv]{1810.04805}~[cs.CL]


\bibitem[duolingo(2023)]%
        {duolingo}
duolingo \bibinfo{year}{2023}\natexlab{}.
\newblock \bibinfo{title}{duolingo}.
\newblock
\newblock
\newblock
\shownote{https://www.duolingo.com/}.


\bibitem[fragment(2022)]%
        {fragment}
fragment \bibinfo{year}{2022}\natexlab{}.
\newblock \bibinfo{title}{fragment}.
\newblock
\newblock
\newblock
\shownote{https://developer.android.com/guide/fragments}.


\bibitem[GUI state(2023)]%
        {GUIstate}
GUI state \bibinfo{year}{2023}\natexlab{}.
\newblock \bibinfo{title}{GUI state}.
\newblock
\newblock
\newblock
\shownote{https://developer.android.com/topic/libraries/architecture/saving-states}.


\bibitem[Gupta et~al\mbox{.}(2021)]%
        {gupta2021layouttransformer}
\bibfield{author}{\bibinfo{person}{Kamal Gupta}, \bibinfo{person}{Justin
  Lazarow}, \bibinfo{person}{Alessandro Achille}, \bibinfo{person}{Larry~S
  Davis}, \bibinfo{person}{Vijay Mahadevan}, {and} \bibinfo{person}{Abhinav
  Shrivastava}.} \bibinfo{year}{2021}\natexlab{}.
\newblock \showarticletitle{Layouttransformer: Layout generation and completion
  with self-attention}. In \bibinfo{booktitle}{\emph{Proceedings of the
  IEEE/CVF International Conference on Computer Vision}}.
  \bibinfo{pages}{1004--1014}.
\newblock


\bibitem[Hochreiter and Schmidhuber(1997)]%
        {hochreiter1997long}
\bibfield{author}{\bibinfo{person}{Sepp Hochreiter} {and}
  \bibinfo{person}{J{\"u}rgen Schmidhuber}.} \bibinfo{year}{1997}\natexlab{}.
\newblock \showarticletitle{Long short-term memory}.
\newblock \bibinfo{journal}{\emph{Neural computation}} \bibinfo{volume}{9},
  \bibinfo{number}{8} (\bibinfo{year}{1997}), \bibinfo{pages}{1735--1780}.
\newblock


\bibitem[Hu and Neamtiu(2011)]%
        {hu2011automating}
\bibfield{author}{\bibinfo{person}{Cuixiong Hu} {and} \bibinfo{person}{Iulian
  Neamtiu}.} \bibinfo{year}{2011}\natexlab{}.
\newblock \showarticletitle{Automating GUI testing for Android applications}.
  In \bibinfo{booktitle}{\emph{Proceedings of the 6th International Workshop on
  Automation of Software Test}}. \bibinfo{pages}{77--83}.
\newblock


\bibitem[Hu et~al\mbox{.}(2019)]%
        {hu2019code}
\bibfield{author}{\bibinfo{person}{Han Hu}, \bibinfo{person}{Qiuyuan Chen},
  {and} \bibinfo{person}{Zhaoyi Liu}.} \bibinfo{year}{2019}\natexlab{}.
\newblock \showarticletitle{Code generation from supervised code embeddings}.
  In \bibinfo{booktitle}{\emph{Neural Information Processing: 26th
  International Conference, ICONIP 2019, Sydney, NSW, Australia, December
  12--15, 2019, Proceedings, Part IV 26}}. Springer, \bibinfo{pages}{388--396}.
\newblock


\bibitem[Hu et~al\mbox{.}(2023a)]%
        {hu2023automated}
\bibfield{author}{\bibinfo{person}{Han Hu}, \bibinfo{person}{Ruiqi Dong},
  \bibinfo{person}{John Grundy}, \bibinfo{person}{Thai~Minh Nguyen},
  \bibinfo{person}{Huaxiao Liu}, {and} \bibinfo{person}{Chunyang Chen}.}
  \bibinfo{year}{2023}\natexlab{a}.
\newblock \bibinfo{title}{Automated Mapping of Adaptive App GUIs from Phones to
  TVs}.
\newblock
\newblock
\showeprint[arxiv]{2307.12522}~[cs.SE]


\bibitem[Hu et~al\mbox{.}(2023b)]%
        {hu2023look}
\bibfield{author}{\bibinfo{person}{Han Hu}, \bibinfo{person}{Yujin Huang},
  \bibinfo{person}{Qiuyuan Chen}, \bibinfo{person}{Terry~Tue Zhuo}, {and}
  \bibinfo{person}{Chunyang Chen}.} \bibinfo{year}{2023}\natexlab{b}.
\newblock \bibinfo{title}{A First Look at On-device Models in iOS Apps}.
\newblock
\newblock
\showeprint[arxiv]{2307.12328}~[cs.SE]


\bibitem[Hu et~al\mbox{.}(2023c)]%
        {hu2023first}
\bibfield{author}{\bibinfo{person}{Han Hu}, \bibinfo{person}{Yujin Huang},
  \bibinfo{person}{Qiuyuan Chen}, \bibinfo{person}{Terry~Yue zhuo}, {and}
  \bibinfo{person}{Chunyang Chen}.} \bibinfo{year}{2023}\natexlab{c}.
\newblock \showarticletitle{A First Look at On-device Models in iOS Apps}.
\newblock \bibinfo{journal}{\emph{ACM Transactions on Software Engineering and
  Methodology}} (\bibinfo{year}{2023}).
\newblock


\bibitem[Hu et~al\mbox{.}(2023d)]%
        {hu2023pairwise}
\bibfield{author}{\bibinfo{person}{Han Hu}, \bibinfo{person}{Haolan Zhan},
  \bibinfo{person}{Yujin Huang}, {and} \bibinfo{person}{Di Liu}.}
  \bibinfo{year}{2023}\natexlab{d}.
\newblock \showarticletitle{Pairwise GUI Dataset Construction Between Android
  Phones and Tablets}.
\newblock \bibinfo{journal}{\emph{arXiv preprint arXiv:2310.04755}}
  (\bibinfo{year}{2023}).
\newblock


\bibitem[Huang et~al\mbox{.}(2021)]%
        {huang2021robustness}
\bibfield{author}{\bibinfo{person}{Yujin Huang}, \bibinfo{person}{Han Hu},
  {and} \bibinfo{person}{Chunyang Chen}.} \bibinfo{year}{2021}\natexlab{}.
\newblock \showarticletitle{Robustness of on-device models: Adversarial attack
  to deep learning models on android apps}. In \bibinfo{booktitle}{\emph{2021
  IEEE/ACM 43rd International Conference on Software Engineering: Software
  Engineering in Practice (ICSE-SEIP)}}. IEEE, \bibinfo{pages}{101--110}.
\newblock


\bibitem[Jyothi et~al\mbox{.}(2019)]%
        {jyothi2019layoutvae}
\bibfield{author}{\bibinfo{person}{Akash~Abdu Jyothi}, \bibinfo{person}{Thibaut
  Durand}, \bibinfo{person}{Jiawei He}, \bibinfo{person}{Leonid Sigal}, {and}
  \bibinfo{person}{Greg Mori}.} \bibinfo{year}{2019}\natexlab{}.
\newblock \showarticletitle{Layoutvae: Stochastic scene layout generation from
  a label set}. In \bibinfo{booktitle}{\emph{Proceedings of the IEEE/CVF
  International Conference on Computer Vision}}. \bibinfo{pages}{9895--9904}.
\newblock


\bibitem[Kingma and Welling(2013)]%
        {kingma2013auto}
\bibfield{author}{\bibinfo{person}{Diederik~P Kingma} {and}
  \bibinfo{person}{Max Welling}.} \bibinfo{year}{2013}\natexlab{}.
\newblock \showarticletitle{Auto-encoding variational bayes}.
\newblock \bibinfo{journal}{\emph{arXiv preprint arXiv:1312.6114}}
  (\bibinfo{year}{2013}).
\newblock


\bibitem[Lee et~al\mbox{.}(2020)]%
        {lee2020guicomp}
\bibfield{author}{\bibinfo{person}{Chunggi Lee}, \bibinfo{person}{Sanghoon
  Kim}, \bibinfo{person}{Dongyun Han}, \bibinfo{person}{Hongjun Yang},
  \bibinfo{person}{Young-Woo Park}, \bibinfo{person}{Bum~Chul Kwon}, {and}
  \bibinfo{person}{Sungahn Ko}.} \bibinfo{year}{2020}\natexlab{}.
\newblock \showarticletitle{GUIComp: A GUI design assistant with real-time,
  multi-faceted feedback}. In \bibinfo{booktitle}{\emph{Proceedings of the 2020
  CHI conference on human factors in computing systems}}.
  \bibinfo{pages}{1--13}.
\newblock


\bibitem[Leiva et~al\mbox{.}(2020)]%
        {leiva2020enrico}
\bibfield{author}{\bibinfo{person}{Luis~A Leiva}, \bibinfo{person}{Asutosh
  Hota}, {and} \bibinfo{person}{Antti Oulasvirta}.}
  \bibinfo{year}{2020}\natexlab{}.
\newblock \showarticletitle{Enrico: A dataset for topic modeling of mobile UI
  designs}. In \bibinfo{booktitle}{\emph{22nd International Conference on
  Human-Computer Interaction with Mobile Devices and Services}}.
  \bibinfo{pages}{1--4}.
\newblock


\bibitem[Li et~al\mbox{.}(2022)]%
        {li2022learning}
\bibfield{author}{\bibinfo{person}{Gang Li}, \bibinfo{person}{Gilles Baechler},
  \bibinfo{person}{Manuel Tragut}, {and} \bibinfo{person}{Yang Li}.}
  \bibinfo{year}{2022}\natexlab{}.
\newblock \showarticletitle{Learning to denoise raw mobile UI layouts for
  improving datasets at scale}. In \bibinfo{booktitle}{\emph{Proceedings of the
  2022 CHI Conference on Human Factors in Computing Systems}}.
  \bibinfo{pages}{1--13}.
\newblock


\bibitem[Li et~al\mbox{.}(2019)]%
        {li2019layoutgan}
\bibfield{author}{\bibinfo{person}{Jianan Li}, \bibinfo{person}{Jimei Yang},
  \bibinfo{person}{Aaron Hertzmann}, \bibinfo{person}{Jianming Zhang}, {and}
  \bibinfo{person}{Tingfa Xu}.} \bibinfo{year}{2019}\natexlab{}.
\newblock \showarticletitle{Layoutgan: Generating graphic layouts with
  wireframe discriminators}.
\newblock \bibinfo{journal}{\emph{arXiv preprint arXiv:1901.06767}}
  (\bibinfo{year}{2019}).
\newblock


\bibitem[Machiry et~al\mbox{.}(2013)]%
        {machiry2013dynodroid}
\bibfield{author}{\bibinfo{person}{Aravind Machiry}, \bibinfo{person}{Rohan
  Tahiliani}, {and} \bibinfo{person}{Mayur Naik}.}
  \bibinfo{year}{2013}\natexlab{}.
\newblock \showarticletitle{Dynodroid: An input generation system for android
  apps}. In \bibinfo{booktitle}{\emph{Proceedings of the 2013 9th Joint Meeting
  on Foundations of Software Engineering}}. \bibinfo{pages}{224--234}.
\newblock


\bibitem[Majeed-Ariss et~al\mbox{.}(2015)]%
        {majeed2015apps}
\bibfield{author}{\bibinfo{person}{Rabiya Majeed-Ariss},
  \bibinfo{person}{Eileen Baildam}, \bibinfo{person}{Malcolm Campbell},
  \bibinfo{person}{Alice Chieng}, \bibinfo{person}{Debbie Fallon},
  \bibinfo{person}{Andrew Hall}, \bibinfo{person}{Janet~E McDonagh},
  \bibinfo{person}{Simon~R Stones}, \bibinfo{person}{Wendy Thomson}, {and}
  \bibinfo{person}{Veronica Swallow}.} \bibinfo{year}{2015}\natexlab{}.
\newblock \showarticletitle{Apps and adolescents: a systematic review of
  adolescents’ use of mobile phone and tablet apps that support personal
  management of their chronic or long-term physical conditions}.
\newblock \bibinfo{journal}{\emph{Journal of medical Internet research}}
  \bibinfo{volume}{17}, \bibinfo{number}{12} (\bibinfo{year}{2015}),
  \bibinfo{pages}{e287}.
\newblock


\bibitem[Memon(2002)]%
        {memon2002gui}
\bibfield{author}{\bibinfo{person}{Atif~M Memon}.}
  \bibinfo{year}{2002}\natexlab{}.
\newblock \showarticletitle{GUI testing: Pitfalls and process}.
\newblock \bibinfo{journal}{\emph{Computer}} \bibinfo{volume}{35},
  \bibinfo{number}{08} (\bibinfo{year}{2002}), \bibinfo{pages}{87--88}.
\newblock


\bibitem[Moran et~al\mbox{.}(2019)]%
        {moran2019redraw}
\bibfield{author}{\bibinfo{person}{Kevin Moran}, \bibinfo{person}{C
  Bernal-Cardenas}, \bibinfo{person}{M Curcio}, \bibinfo{person}{R Bonett},
  {and} \bibinfo{person}{D Poshyvanyk}.} \bibinfo{year}{2019}\natexlab{}.
\newblock \bibinfo{title}{The ReDraw Dataset: A Set of Android Screenshots, GUI
  Metadata, and Labeled Images of GUI Components}.
\newblock
\newblock


\bibitem[Oulasvirta et~al\mbox{.}(2020)]%
        {oulasvirta2020combinatorial}
\bibfield{author}{\bibinfo{person}{Antti Oulasvirta},
  \bibinfo{person}{Niraj~Ramesh Dayama}, \bibinfo{person}{Morteza Shiripour},
  \bibinfo{person}{Maximilian John}, {and} \bibinfo{person}{Andreas
  Karrenbauer}.} \bibinfo{year}{2020}\natexlab{}.
\newblock \showarticletitle{Combinatorial optimization of graphical user
  interface designs}.
\newblock \bibinfo{journal}{\emph{Proc. IEEE}} \bibinfo{volume}{108},
  \bibinfo{number}{3} (\bibinfo{year}{2020}), \bibinfo{pages}{434--464}.
\newblock


\bibitem[spotify(2023)]%
        {spotify}
spotify \bibinfo{year}{2023}\natexlab{}.
\newblock \bibinfo{title}{spotify}.
\newblock
\newblock
\newblock
\shownote{https://en.wikipedia.org/wiki/Spotify}.


\bibitem[supportDifScr(2023)]%
        {supportDifScr}
supportDifScr \bibinfo{year}{2023}\natexlab{}.
\newblock \bibinfo{title}{supportDifScr}.
\newblock
\newblock
\newblock
\shownote{https://developer.android.com/guide/topics/large-screens/support-different-screen-sizes}.


\bibitem[tabletShare(2023)]%
        {tabletShare}
tabletShare \bibinfo{year}{2023}\natexlab{}.
\newblock \bibinfo{title}{tabletShare}.
\newblock
\newblock
\newblock
\shownote{https://gs.statcounter.com/platform-market-share/desktop-mobile-tablet}.


\bibitem[Theis et~al\mbox{.}(2015)]%
        {theis2015note}
\bibfield{author}{\bibinfo{person}{Lucas Theis}, \bibinfo{person}{A{\"a}ron
  van~den Oord}, {and} \bibinfo{person}{Matthias Bethge}.}
  \bibinfo{year}{2015}\natexlab{}.
\newblock \showarticletitle{A note on the evaluation of generative models}.
\newblock \bibinfo{journal}{\emph{arXiv preprint arXiv:1511.01844}}
  (\bibinfo{year}{2015}).
\newblock


\bibitem[uiautomator2(2023)]%
        {uiautomator2}
uiautomator2 \bibinfo{year}{2023}\natexlab{}.
\newblock \bibinfo{title}{uiautomator2}.
\newblock
\newblock
\newblock
\shownote{https://play.google.com/store}.


\bibitem[Vaswani et~al\mbox{.}(2017)]%
        {vaswani2017attention}
\bibfield{author}{\bibinfo{person}{Ashish Vaswani}, \bibinfo{person}{Noam
  Shazeer}, \bibinfo{person}{Niki Parmar}, \bibinfo{person}{Jakob Uszkoreit},
  \bibinfo{person}{Llion Jones}, \bibinfo{person}{Aidan~N Gomez},
  \bibinfo{person}{{\L}ukasz Kaiser}, {and} \bibinfo{person}{Illia
  Polosukhin}.} \bibinfo{year}{2017}\natexlab{}.
\newblock \showarticletitle{Attention is all you need}.
\newblock \bibinfo{journal}{\emph{Advances in neural information processing
  systems}}  \bibinfo{volume}{30} (\bibinfo{year}{2017}).
\newblock


\bibitem[Wang et~al\mbox{.}(2021)]%
        {wang2021screen2words}
\bibfield{author}{\bibinfo{person}{Bryan Wang}, \bibinfo{person}{Gang Li},
  \bibinfo{person}{Xin Zhou}, \bibinfo{person}{Zhourong Chen},
  \bibinfo{person}{Tovi Grossman}, {and} \bibinfo{person}{Yang Li}.}
  \bibinfo{year}{2021}\natexlab{}.
\newblock \showarticletitle{Screen2words: Automatic mobile UI summarization
  with multimodal learning}. In \bibinfo{booktitle}{\emph{The 34th Annual ACM
  Symposium on User Interface Software and Technology}}.
  \bibinfo{pages}{498--510}.
\newblock


\bibitem[youtube(2023)]%
        {youtube}
youtube \bibinfo{year}{2023}\natexlab{}.
\newblock \bibinfo{title}{youtube}.
\newblock
\newblock
\newblock
\shownote{https://en.wikipedia.org/wiki/YouTube}.


\bibitem[Zhao et~al\mbox{.}(2022)]%
        {zhao2022code}
\bibfield{author}{\bibinfo{person}{Yanjie Zhao}, \bibinfo{person}{Li Li},
  \bibinfo{person}{Xiaoyu Sun}, \bibinfo{person}{Pei Liu}, {and}
  \bibinfo{person}{John Grundy}.} \bibinfo{year}{2022}\natexlab{}.
\newblock \showarticletitle{Code implementation recommendation for Android GUI
  components}. In \bibinfo{booktitle}{\emph{Proceedings of the ACM/IEEE 44th
  International Conference on Software Engineering: Companion Proceedings}}.
  \bibinfo{pages}{31--35}.
\newblock


\end{thebibliography}

\end{document}